# A Survey of P2P multidimensional indexing structures

Ewout M. Bongers, Johan Pouwelse

TU Delft – Course IN4306 – Literature Survey

*Abstract* — **Traditional databases have long since reaped the benefits of multidimensional indexes. Numerous proposals in the literature describe multidimensional index designs for P2P systems. However, none of these designs have had real world implementations. Several proposals for P2P multidimensional indexes are reviewed and analyzed. Znet and VBI-tree are the most promising from a technical standpoint. All of the proposed designs assume honest nodes and are thus open to abuse. This is a critical flaw that must be solved before any of the proposed systems can be used.**

## I. Introduction

Since their inception P2P networks have included some form of indexing to locate items of interest. Such as DHTs that enable keyword search queries. Queries for ranges of keys have proven to be slightly more challenging for P2P. But the holy grail of indexing (in general) is supporting a multidimensional range query and this is more challenging still to realize in P2P.

To illustrate the usefulness of multidimensional range queries, imagine wanting to search a P2P network based on geo-locations (figure 1). This is infeasible without an index. But even having an index over the individual dimensions is not enough. With such indexes the red and green query regions in the figure would need to be queried and then intersected, a very costly and inefficient operation. This problem grows when dimensions are added. Contrary to this, a true multidimensional index not only searches all dimensions in one query, but even gets more selective as dimensions are added.

In singly system situations, database and games engineers have devised a plethora of indexing structures for multidimensional data. These are used for everything from enterprise data warehouses to collision detection in 3D. However these structures are not suited to use in a P2P environment, if not downright impossible. Supporting these multidimensional range queries has proven to be challenging for P2P networks. Therefor this survey attempts to answer the question: *What is the state of the art for multidimensional range queries in P2P networks?*

The organization of this literature survey is as follows. Section II shows prominent indexing structures as found in P2P literature. Skip lists and skip graphs are briefly discussed, as they are central to many of the index structures. Section III analyzes the strengths and weaknesses of the different indexing structures. Section IV looks at security issues surrounding the indexing structures.

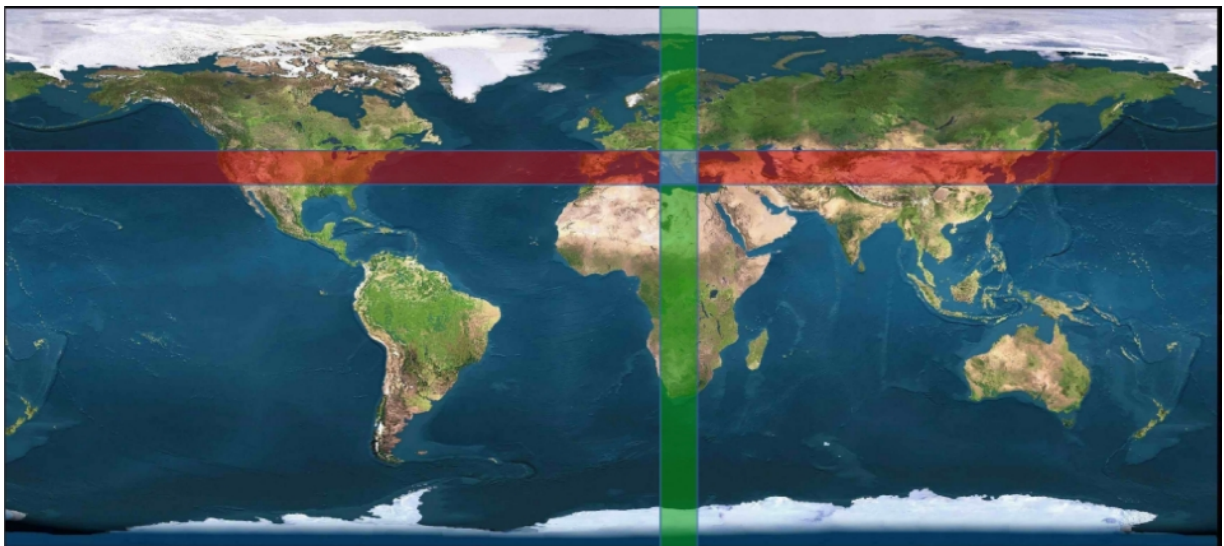

*Figure 1: Intersecting latitude and longitude ranges*

II. MULTIDIMENSIONAL INDEXING STRUCTURES

The methods devised for multidimensional indexing (summarized in table 1) roughly fall into two basic categories, first is reducing the dimensionality, the second is partitioning the multidimensional space. Reducing dimensionality is usually done with a space filling curve, a curve that sequentially numbers all the discrete points of multidimensional space. This single dimensional space is then stored using conventional methods like DHTs or skip graphs. Space partitioning is recursively subdividing space to yield regions that contain a relatively small number of keys. As will be shown there are all manner of varieties that mix and combine partitioning with reduction.

Some indexing structures are based on the well known DHT, but others are based on the lesser known skip graph, which is in turn based on the skip list. For the benefit of the lesser informed reader, the skip list and graph will be presented first.

A. Skip Lists

In 1990 Pugh [1] introduced Skip Lists. The data structure starts as a normal linked list (level 0). With a fixed probability $p$ each element in the list also participates in a higher level linked list (level 1, 2, 3, ...). By choosing a suitable $p$, the expectation is that the highest level is populated with a few nodes, and can thus "skip" through the lowest level list. The structure has (withing a certain margin of error) characteristics that are very similar to B+ trees (where the parameter $p$ determines the fan out). However

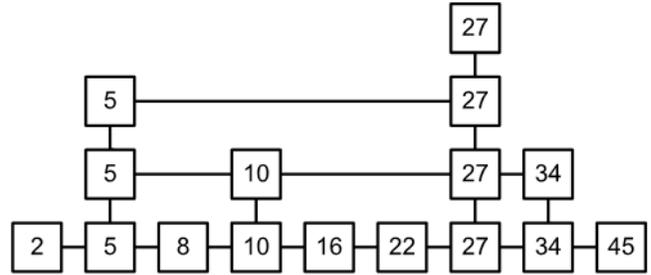
Figure 2: A skip list [5]

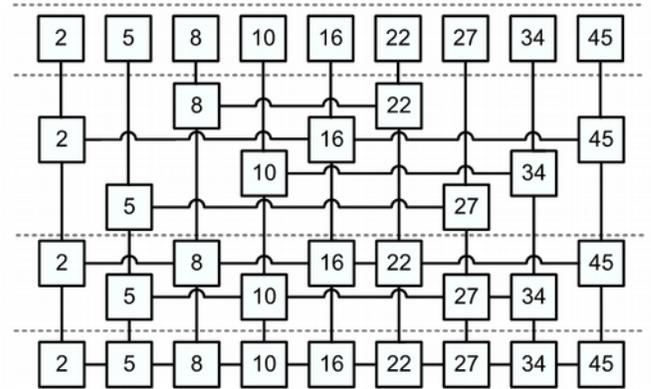
Figure 3: A skip graph [5]

insertion and deletion are much simpler.

Searching begins with the highest level and descends down if the next key in the linked list is larger than the search key. Eventually on level 0, the search key can be found (or the start of a range). The number of steps needed to find a key is on average O(log n), with the log base a function of $p$. Insertion is as simple as inserting an element in the level 0 list, and randomizing if it should be included in the next level, and link it in if needed. Deletion requires that the element be unlinked from all lists it is included in. The skip list also has very favorable concurrency characteristics.

B. Skip Graphs

Aspnes and Shah [2], adapted the concept of skip lists to peer to peer systems in the form of a skip graph.

|  | Node Scaling Performance | | Dimension Scaling Performance | | Load Balance | Robust | Security | Drawback |
|---|---|---|---|---|---|---|---|---|
|  | Messaging | Storage | Messaging | Storage |  |  |  |  |
| Skip List | O(log i) | O(i) | - | - | - | - | - | Not P2P |
| Skip Graph | O(log n) | O(log n) | - | - | - | - | - | Single Dimensional |
| SCRAP | O(log n) | O(log n) | ? | O(1) | Passive | - | - | Very poor locality |
| CISS | O(log n) | O(log n) | O(1) | O(1) | Active | ± | - | Poor locality |
| Squid | O(log n) | O(log n) | O(1) | O(1) | Passive+Active | ± | - | Poor locality |
| MAAN | O(log n) | O(log n) | O(1) | O(k) | Passive | ± | - | Depend on selectivity |
| Mercury | O(log n) | O(log n) | O(1) | O(k) | Passive | ± | - | Depend on selectivity |
| MURK | O(log n) | ? | ? | $O(k^c)$ | Passive | - | - | Storage Cost & Balance |
| Znet | O(log n) | O(log n) | O(1) | O(k) | Passive+Active | ± | - | Poor locality |
| SkipIndex | O(log n) | O(log n) | O(1) | O(k) | Active | + | - | Poor locality |
| VBI+R-Tree | O(log n) | O(log n) | ? | O(1) | Active | + | - | Congestion & Complex |

Table 1: Comparison of multidimensional P2P indexing structures (i: data items, n: network nodes, k: dimensions)

Many multidimensional query solutions are based on this concept of skip graphs.

The level 0 is the same as the regular skip list, it is an ordered list of elements (nodes in the case of a P2P overlay). For level 1 instead of randomizing a join-or-not value, each element randomizes what level 1 list to join. Since most P2P systems use randomized node id's, Aspnes and Shah suggest to use the node ID for these random values. If level 1 is consists of $n$ lists, the first $log(n)$ bits of the random value should be used to determine what level 1 list to join. For the next level the first $2log(n)$ random bits decide what list to join. This continues until the empty list is reached. Notice that each list is identified by a bit string. All elements in a list share that lists identification bit string as prefix of their own random IDs.

The skip graph structure provides a P2P overlay structure that works as an ordered list. It is important to note that the simple ordinal operations $<$ and $>$ are only defined on a single dimension. For example the tuple (1, 2) is not greater than nor smaller than the tuple (2, 1). As such Skip Graphs only work for creating single dimensional structures and only support single dimensional range queries, though this is still an advancement over the hashed DHT structures. Also the Skip Graph structure does not expressly impose a method for actually ordering elements, only which elements/nodes join what lists.

A great advantage of the Skip Graphs over other P2P ordered index structures is that it makes no a priori assumptions about the range of the dimensions. As will be shown, some indexing structures are based on knowing the range of values that a dimension spans, whereas the Skip Graph only requires that the dimension on which it works supports ordinal operators.

Aspnes and Shah hint at multidimensional data as a possible future topic of research. However it appears that this avenue of research was not pursued.

### C. Space Filling Curves

Others did pick up the adaptation of Skip Graphs to multidimensional data. By adapting techniques from the databases field, SCRAP [3] employs a z-ordering space filling curve to linearize the multidimensional search space. This linearized space is then stored in a Skip Graph where each node is responsible for a certain range in the z-curve.

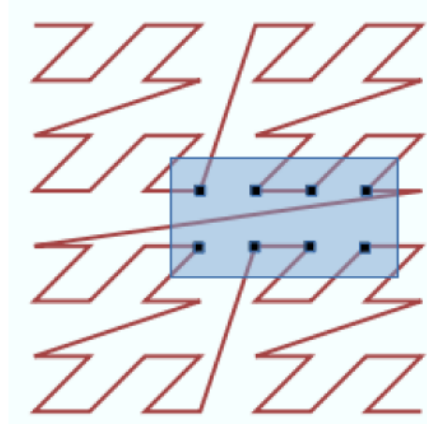

Figure 4: Intersecting a space filling curve

SCRAP assumes that the dimensions are z-curve linearizable. This means interleaving the binary representation of values from each dimension. Thus the $<x, y>$ tuple $<0100, 0101>$ would be mapped to value $00110001$. This method, known as z-ordering, works for most common binary encodings of integers and floating point numbers. By selecting specific ranges on the linearized space, it is possible to perform multidimensional range queries.

However there are several downsides to this linearisation. For one, it can only support numerical values. Second, it limits the input numbers to a more or less similar domain, or phrased differently, a similar number of significant bits. Third, one dimension is "dominant". By the nature of the interleaving of bits, one dimension will provide the most significant bit of the linearized space. Similarly, the second dimension will be dominant over a third. Any query that spans the range of a dominant dimension, will need to be split into multiple smaller queries, and thus might not efficiently map to the division of the z-ordering.

This last drawback is one that is generally applicable to space filling curves and is also suffered by traditional database systems that employ space filling curves. Selecting a region can intersect the curve many times [7]. For example consider figure 4. The z-ordering in this figure covers an 8 * 8 grid. Thus the space-filling curve visits all the 64 points that comprise this space. However selecting exactly the shaded region involves 6 distinct segments of the z-curve's linear space. Thus multidimensional range queries have to be split up into many range queries on the linearized single dimensional index. Another possibility is to query slightly coarser sections and filter any false positives after the query. The trade-off here is post-processing. If the split is too coarse, many false positives will need to be filtered out. Too fine a split leads to an explosion of smaller range queries, but reduces the number of false positives. Also calculating the intersection points of the space filling curve with the query range can be simple but is not necessarily trivial, so a fine split could require significant processing power to achieve. Lastly, when the dimensionality goes up, the false positives will more frequently be located on nodes unrelated to the original query, thus involving a larger portion of the network in a query. Or interpreted differently, higher dimensional space filling curves have poor locality.

The z-curve is relatively simple among space-filling curves. Generally a Hilbert-curve will offer better locality, by avoiding long jumps of the curve. Hilbert curves even provide a bound on distance between two points on the curve relative to the euclidean distance [8]. So it is only natural that some systems use Hilbert-curves.

Both CISS [23] and Squid [25] use a Hilbert-curve with the express intention of maintaining locality in the linearized space. Functionally CISS and Squid are similar to SCRAP except they do not store keys in a Skip Graph but rather use DHTs. This immediately raises the question of load balancing, since the linearized space is not filled uniformly. This is where CISS and Squid are a little different. CISS specifies a local and global strategy to reorder nodes along the DHT ring such that the load on each node is approximately even. Squid uses a single more expensive algorithm to distribute nodes along the DHT. Squid also specifies a mechanism for distributing node load as they join the DHT.

Be aware that as noted in [3], any space-filling curve based system will be afflicted by the curse of dimensionality. The curse of dimensionality refers to the exponential increase of "space" as dimensions are added. The result of which is that *all* data points are "very far" relative to each other because the space is so empty. Effectively, the concepts of euclidean distance and locality lose their meaning. Therefor SCRAP, CISS and Squid will not scale particularly well in the number of dimensions, nor will any system based on space-filling curves.

### D. Locality Preserving Hashing

Very similar to space filling curves is another strategy to reduce the multidimensional aspect of complex data to a single dimensional problem. By using locality preserving hashes the Multi-Attribute Addressable Network (MAAN) [4] aims to do just that.

With locality preserving hashing, dimensions are directly mapped to a linear space. MAAN chooses to use a Chord DHTs as its linearized space. The underlying Chord DHT can then be used to look up ranges in each dimensions.

To produce a uniform output of the hashing function, the (cumulative) distribution function for the values along each dimension (or *attribute* in MAAN) must be known at design time. By inverting the distribution function, the output space of a locality preserving hash function is uniformly filled and ordering (and thus locality) is preserved. When failing to provide an accurate distribution function the load on the DHT nodes can become skewed.

Queries on a given dimension are executed by sending a search request to the *successor($H(v_l)$)* where $H$ is the hash function and $v_l$ the lower bound of the search. The search then continues along the DHT ring until the upper bound is passed. Each node adds to the results as the search request passes it.

By knowing the distribution function for each dimension, it is possible to determine the minimum and maximum values for each dimension. This knowledge allows MAAN to estimate the selectivity of a query on each dimension. It uses this to reduce message complexity during query operations. By choosing to query the dimension with the smallest selectivity, or put differently, the dimension who's query range spans the shortest section on the DHT, the number of nodes visited for a query is minimized. The query ranges of the other dimensions are sent along with the query as filters for the values found along the way. This avoids sending around unneeded search results that would be filtered out in the end. Notice that if the query range for every dimension has poor selectivity then MAAN can walk a significant part of the DHT, even if the combination of range queries is highly selective.

The a priori knowledge of dimension minimum and maximum make MAAN unattractive as a general solution to multidimensional indexing in P2P systems. Even if this information is known at design time, it cannot evolve with the system. Furthermore, to ensure balanced performance one would also have to know the distribution functions of the dimensions at design time.

For general applications it is unlikely that these parameters will not change over time.

The only performance metric of MAAN that has received attention is the routing cost, and in that respect MAAN preforms well. The routing cost is independent of dimensionality, since only one primary dimension is queried. However considering that a query travels linearly along a portion of the DHT ring, and that each node has to do non trivial processing before passing along the query, it would seem that response time to a query can be high, making MAAN unattractive for interactive usage. Another performance metric where MAAN does not shine is storage cost. Each item to be indexed must be stored in the DHT multiple times, once for each dimension. Thus storage cost is linear to dimensionality.

Very similar to MAAN is Mercury [20]. It too uses a DHT ring for each dimension and it has properties that are nearly identical to MAAN. However, Mercury uses random sampling of the key spaces to dynamically approximate a distribution function for each dimension. The cost for this advantage is more algorithms and algorithms of greater complexity.

### E. Space Partitioning

Instead of simplifying high dimensional space into a simpler representation, it is also possible to exploit the distribution of data and deal with the multidimensional space directly. This class of indexing structures is known as space partitioning indexes and they function by iteratively subdividing space to smaller spaces. Notable examples from the databases domain are kD-trees [9], octrees [10] and R-trees [13].

MURK [3] applies a kD-tree to the multidimensional data. A kD-tree recursively subdivides a space based on the median of the values inserted in one of the dimensions. Thus each split is guaranteed to separate half the data points. Usually it will cycle through the dimensions when splitting, the first level splitting on the first dimension, the second level splitting on the second dimension, and so on. In the P2P version of the kD-tree proposed by MURK each node is made responsible for one leaf of the kD-tree, and the data therein. If a leaf is full, it can be split and the data will be redistributed over two nodes. Thus the tree forms along with the data distribution, and distributes the load fairly. As noted in [3], this structure bears a strong resemblance to the CAN [22] network overlay. However in CAN the hash output space is split, not the data. Although it could be argued that since CAN uses hashes that produce uniform output, splitting the hash output space is equally effective as splitting the data space directly.

Compared to the space linearizing strategies, MURK requires more complex algorithms for joining, failing or leaving nodes. Also kD-trees are not inherently balanced. Therefor using them in a dynamic situation is challenging. MURK is not capable of tree balancing operations. However simulation results show that it performs reasonably well with non-uniform/skewed data sets.

Much like CAN, MURK cannot bound the number of neighbor nodes it has to track. The neighbor along one region boundary might be subdivided many times, leading to many neighbor pointers. In practice it seems that the average number of neighbors is on the order of $O(k)$ to $O(k^2)$.

### F. Znet

Other space partitionings are possible too. Znet [21] is based on the realization that a z-curve is a form of space partitioning. If for example the value of three dimensional z-curve starts with the binary prefix 001, this means that the value is in the lower half of the first and second dimension but in the upper half of the third. At the next level, the next 3 bits of the prefix provide further positioning within the space defined by the prefix. This space partitioning is equivalent to a regular octree but has a total ordering. A 2D equivalent of this can be seen in figure 4 where the MSB of the curve would indicate top or bottom half and the next bit would indicate left or right half of that top or bottom half.

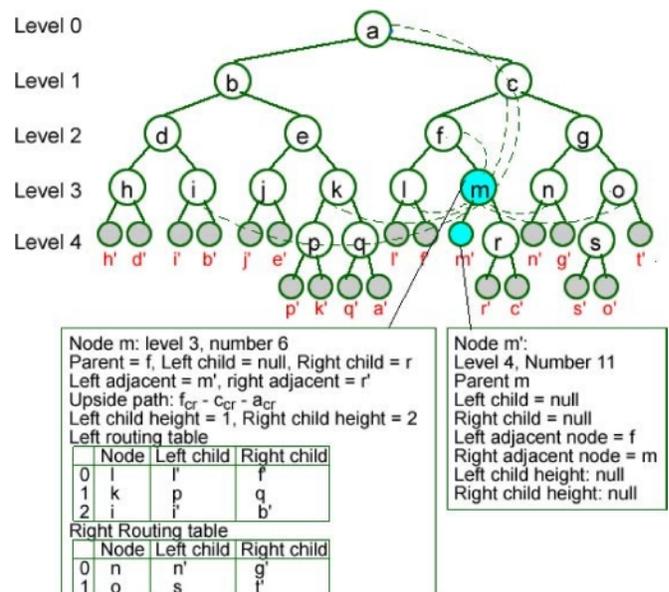

Figure 5: VBI-tree structure[12]

The cleverness of Znet is in mapping the partitionings to P2P nodes. Instead of directly storing a z-ordered space in a skip graph, it forms a skip graph of z-order prefixes. Each node is assigned such a prefix and is linked into the skip graph. The advantage of this is that each node is responsible for a rectangular section of the index space. This simplifies message routing for range queries that are axis aligned. Using this space partitioning, Znet shows it is possible to dynamically partition space for a skewed data set by partitioning to a level where each node has a similar load.

Another indexing structure that is very similar to (and proposed at almost the same time as) Znet is SkipIndex [11]. It too relies on a total ordering of partitioned space. However, like MURK, it achieves this by splitting directly on the data, not on a z-ordering. This leads to a kD-tree like space partitioning as opposed to a regularly sized space partitioning in Znet. Thus merging nodes in SkipIndex is a little more complicated than Znet. Also, compared to Znet, the query routing is less efficient in SkipIndex.

### G. VBI-tree

A more abstract way of space partitioning is the ability to map tree structures to P2P networks. With this ability, the traditional space partitioning algorithms of databases can be used. This is the aim of the VBI-tree [12]. It introduces a framework to map well-known multidimensional index tree structures to P2P networks. This enables many well-known space partitioning trees (such as R(*)-trees [13], M-trees [14], X-trees [6], etc.) to be deployed in a P2P network. The obvious advantage being that the technique of multidimensional indexing is separated from the intricacies of P2P networking.

The VBI-tree is a slight adaptation of the BAlanced Tree Overlay Network (BATON [15]). In the VBI-tree nodes are ordered in a tree structure, and each node keeps parent and child links. The nodes also keep left and right sibling links such that each next sibling link points to a node twice as distant as the node that the previous sibling link pointed too (in the example figure 5, $m$ points to $l$ and the twice as far $k$, and again twice as far $i$). Each link to a sibling node also includes pointers to the left and right children of this sibling node. This enables recovery from node failure but also serves to distribute information about the fullness of each level.

As long as the chosen multidimensional indexing algorithm results in a balanced tree, as indeed R-, M- and X-trees do, then each VBI node can directly represent a node from these trees. If the underlying algorithm requires it, common tree restructuring operations are also supported on VBI-tree to enable the tree to remain balanced.

The main disadvantage of the VBI structure is that it forces the underlying algorithm to use a binary split. Put differently, a fan-out factor of 2. This might not be an optimal or ideal condition for all underlying algorithms and data. Another practical issue is that many traditional tree algorithms require locking during updates, it is unclear how this would be supported in VBI. Lastly note that in many tree algorithms the root node is a bottleneck for updates, VBI cannot avoid this congestion.

### III. ANALYSIS

The reviewed P2P indexing structures that support multidimensional range queries will be evaluated on the following properties:

- Algorithmic complexity: with respect to the number of nodes in the network and the number of dimensions;
- Load balance: does the network distribute the load evenly;
- Robustness: dealing with node failures;
- Security: dealing with corruption or active attack;
- Practical considerations: what problems the practical implementations might encounter.

And the following symbols will be used

- $n$, the number of nodes
- $k$, the number of dimensions
- $i$, the number of keys in the index

Refer to table 1 for a summary of the main properties of each indexing structure.

## A. Algorithmic Complexity

With respect to the number of nodes in the network, all indexing structures claim $O(\log n)$ messaging complexity with respect to network size. Note that most indexing structures actually have $O(\log n + n*s)$ where $s$ is query selectivity, thus a linear fraction of n is involved. However it is assumed that $s$ will be small in practical situations, so this term should be almost constant. Therefor, interpret the messaging complexity with care. MAAN and Mercury are particularly sneaky in this respect. They actively pick the dimension with the smallest $s$, but the selectivity of the query as a whole might be orders of magnitude smaller. Leading to a relatively large $n*s$ component.

With the exception of VBI-tree, the indexing structures use either a DHT or a Skip Graph to store their index in the P2P network. Both have O(log n) storage complexity with respect to the network size. VBI tree has an almost constant sized state, except for the on-level routing table, which ends up making it $O(\log n)$ also.

## B. Dimensional Message Complexity

All indexing structures that employ some form of space linearisation are invariant to the number of dimensions, and thus have dimensional messaging complexity $O(1)$. So for a query, relevant nodes are reached in $O(\log n)$ hops no matter what the number of dimensions is. The tradeoff for this is poor locality. With high dimensional spaces, the locality of space filling curves is lost. Making a query will effectively flood the query message on the network, in $O(\log n)$ steps. With the exception of [3] all indexing structures do not to draw attention to this in their respective results sections.

The kD-tree like structure of MURK is hard to express in terms of dimensionality. However, assuming uniform data, each division still produces two equal halves, so given the same number of key values, the structure should still use the same number of splits. For messaging complexity an increasing dimensionality does mean there are more neighbors to route to for each leaf node. This reduces the messaging complexity to something in the realm of $O(1/k^c)$.

VBI-tree is obviously somewhat dependent on the characteristics of the space partitioning tree that is mapped to the P2P overlay. However since it is a binary split, much like the kD-tree, the number of splits should again be similar, irrespective of dimensionality. Since VBI is defined for balanced trees, it's structure as P2P overlay is thus similar, if not the same, for every dimensionality. So are the routing tables, and thus dimensional messaging complexity is on the order of $O(1)$. Note however that similar to space linearisation losing locality in higher dimensional spaces, so can the mapped space partitioning tree. For example when used as an R-Tree. With $k >= 5$ the nodes of an R-trees approach 90% overlap [6], this degrades the discriminatory value of each node and leads to the network more or less flooding the query. However other structures such as the X-tree effectively avoid this problem, but are complex to implement.

## C. Dimensional Storage Complexity

The dimensional storage complexity is more varied. Again the structures that use space linearisation are mostly invariant to the dimensionality, but the others (MAAN, Mercury and MURK) compare unfavorably. MAAN and Mercury need to store each value once for each dimension, so they scale with O(k). With increasing dimensions, MURK has an increasing number of neighbors for each leaf node, Interpreted from comparative results graphs in literature, it appears that MURK storage scales as $O(k^c)$, with $c$ on the order of 1.5 to 3. Again VBI is the odd one out, in that it is relatively stable to changes in the number of dimensions.

## D. Load Balance

Each indexing structure claims to result in a balanced system. However the semantics of this claim vary somewhat between the indexing structures. On the one hand there is keeping the query load balanced, but this is not the same as keeping the storage load balanced, which is different again to keeping the messaging load balanced.

For balancing these loads, different mechanisms are employed. They seem to fall in roughly two categories. Firstly there is passive balancing in which regions/keyspace from leaving nodes are taken over by other nodes and "unbalanced" regions or nodes are split off to new arrivals. This scheme depends on network churn for balancing. This fails to address situations where network churn is not high enough to keep up with the natural unbalancing processes. Secondly there is active balancing. In this case nodes are continually shifting data too one another in an effort to distribute load fairly. This is based on algorithms to detect or estimate load. So adding some complexity to the indexing structure. The passive and active methods can also be combined as in Znet. The Skip Graph based methods can also use the system in [24] to actively balance themselves.

For all the attention on storage balance, there is little focus on balancing query load. In fact the implicit assumption seems to be that index keys are queried more or less uniformly and thus query balance would be achieved through balancing storage. There is no argumentation supporting this assumption. Although in most situations it would be reasonable to assume that a simple caching strategy can alleviate the worst of this, but again there is no supporting argumentation for this.

### E. Robustness

Since link and node failures are common in P2P networking, any overlay should be designed to deal with this. However SCRAP and MURK do not present anything in this regard. CISS, Squid, MAAN and Mercury rely on the underlying DHT to provide redundancy. DHTs have proven to work in real world so this approach is feasible, although it would be interesting to see an analysis on the effort needed to break one of the DHTs, especially in the case of MAAN and Mercury. Since they use one DHT per dimension, they are not sized to the full network, and might thus be more vulnerable than larger DHTs.

Znet, SkipIndex and VBI-tree each have redundancy built in. Znet is rather simplistic, just like a DHT each node in the underlying Skip Graph also maintains the state of a fixed number of predecessors. DHTs have proven this can work, so it is reasonable to expect similar results when used in a Skip Graph. A little less clear is how to deal with a broken link in a Skip Graph, this is not made explicit in the literature.

The skip graph based approaches (MURK, Znet, SkipIndex) can use a more robust version of skip graphs called a Rainbow Skip Graph [5] instead of regular skip graphs. The rainbow skip graph is especially designed to provide a high degree of redundancy and also manages to reduce the number of neighbor connections to $O(1)$.

The VBI-tree indexing structure detects broken links, reroutes queries and starts a restructuring process. However, this is only half the story. VBI-tree seems to have no data redundancy, even though it could be added. Without it VBI-tree is not usable in practical applications.

### F. Concurrency

The VBI-tree network overlay maps any balanced tree structure to a P2P network. However there are many such tree algorithms to choose, and most of these require locking of some sort. Typically in balanced trees, node splits happen in the leaves and splits continue upwards to the root as long as the parent nodes are full. After a child has split, the parent needs to decide to split or not, and during this time the portion of the child that is split off is not reachable. Thus the data structure is inconsistent if the parent does not lock during the update of the child. This chain of locks can continue all the way to the root.

An obvious solution would be to use a lock-free balanced space partitioning tree, however no tree with this property appears to exist in literature. Thus VBI-tree needs the distributed system to use exclusive locks, an unattractive quality in any distributed system, since it will easily lead to congestion. So barring any breakthrough discovery in lock-free space partitioning trees, VBI-tree is not a practical option.

## IV. SECURITY

Although the indexing structures give consideration to natural faults and disruption, and actively replicate and repair, none of them can deal with active attack, disruption and pollution. This makes them too fragile to deploy in real world situations. Specifically, the indexing structures do nothing to mitigate the inherent weaknesses of the underlying DHTs and Skip Graphs.

The standard DHT has many known attacks and even more defenses [16]. An important one is the Sybil attack. In this attack an adversary tries to insert a large number of malicious nodes into the network. With these nodes the attacker can disrupt the network, pollute the indexes or carry out more targeted attacks. A malicious node in an indexing structure can misdirect queries, manipulate them, give false answers or even selectively drop certain queries. The possibilities are endless.

It is known that Sybil attacks are possible for DHTs,

and it's not hard to imagine them working in Skip Graphs as well. The Sybil attack usually exploits the fact that, for scalability, single nodes only have a very limited view of the whole network. It is very hard to defend against a Sybil attack with only this limited knowledge. Skip Graphs rely on a similar limited view of the whole network and thus should be vulnerable to a Sybil attack. Without a defense, all presented indexing structures based on Skip Graphs are vulnerable to renting a cheap VPS and having it disrupt the network.

A little more sophisticated is the Eclipse attack. In this scenario an attacker manages to become all, or a large fraction of, the neighbors of a node. This enables the attacker to control the flow of information to and from the eclipsed node. Effectively the eclipse attack is a targeted form of the Sybil attack. Because of its smaller scale and targeted nature, an eclipse attack is harder to detect than a Sybil attack.

Performing an Eclipse attack on a Skip Graph is not as simple as attacking a simple DHT. When the membership vector of the different levels of the Skip Graph is determined by a hash of verifiable data (such as public IP address) then an attacker would need access to a vast part of the IP address space to fully eclipse a node. Not to mention the computation required to hash all the IP addresses. Moreover, when two nodes are neighbors in one level they are, with high probability, not neighbors in higher or lower levels. Thus every level of a node would need to be infiltrated to eclipse a node fully, with a simple IP hash defense, this quickly becomes impractical. Note that IPv6 slightly weakens this scheme, since computers have more freedom to select an IP address, but that still leaves a computational problem.

## V. Conclusion

None of the described indexing structures are usable yet. Issues such as security, locking and redundancy need to be addressed before any of the indexing structures can be considered for practical applications.

From a technical standpoint, Znet, MURK and VBI-tree are the indexing structures that do not disqualify themselves outright. MURK is hard to analyze theoretically, therefor it needs to prove itself in real world applications. As of writing this has not happened. Znet suffers a little from reduced locality in higher dimensions because it is a space linearisation method. In combination with rainbow skip graphs for fault tolerance, Znet could very well be useful. Lastly VBI-tree is complex and its usefulness depends strongly on the space partitioning algorithm that is mapped to the P2P overlay.